\documentclass[aps,twocolumn,epsfig,prb]{revtex4}

\newcommand{\bsigma}{\mbox{\boldmath $\sigma$}}

\usepackage{amsmath}
\usepackage{graphicx}

\def\nn{\nonumber}

\begin{document}

\title{Soliton Trap in Strained Graphene Nanoribbons}

\author{Ken-ichi Sasaki}
\email[Email address: ]{SASAKI.Kenichi@nims.go.jp}
\affiliation{International Center for Materials Nanoarchitectonics, 
National Institute for Materials Science,
Namiki, Tsukuba 305-0044, Japan}

\author{Riichiro Saito}
\affiliation{Department of Physics, Tohoku University, Sendai 980-8578,
Japan}

\author{Mildred S. Dresselhaus}
\affiliation{Department of Physics, Department of Electrical Engineering
and Computer Science, Massachusetts Institute of Technology, Cambridge,
MA 02139-4307}

\author{Katsunori Wakabayashi}
\affiliation{International Center for Materials Nanoarchitectonics, 
National Institute for Materials Science,
Namiki, Tsukuba 305-0044, Japan}
\affiliation{PRESTO, Japan Science and Technology Agency,
Kawaguchi 332-0012, Japan}

\author{Toshiaki Enoki}
\affiliation{Department of Chemistry, Tokyo Institute of Technology,
Ookayama, Meguro-ku, Tokyo 152-8551, Japan}

\date{\today}

\begin{abstract}
The wavefunction of a massless fermion 
consists of two chiralities, left-handed and right-handed,
which are eigenstates of the chiral operator.
The theory of weak interactions of elementally particle physics
is not symmetric about the two chiralities, and 
such a symmetry breaking theory is referred to as a chiral gauge theory.
The chiral gauge theory can be applied
to the massless Dirac particles of graphene. 
In this paper we show 
within the framework of the chiral gauge theory for graphene
that a topological soliton exists near the boundary of a graphene
nanoribbon in the presence of a strain.
This soliton is a zero-energy state connecting two chiralities 
and is an elementally excitation transporting a pseudospin. 
The soliton should be observable by means of 
a scanning tunneling microscopy experiment. 
\end{abstract}

\maketitle

For a massless fermion, 
the left-handed and right-handed chiralities 
are a good quantum number and 
the two chirality eigenstates evolve independently 
according to the Weyl equations. 
One chirality state goes into the other chirality state 
under a change in parity.
The weak interactions in elementary particle physics
act differently on the left-handed and right-handed states,
which results in well-known phenomena, 
such as the parity violation
for nuclear $\beta$ decay.~\cite{sakurai67}
The weak force is described by a gauge field. 
In general, 
a gauge field which has a different (the same) sign of the coupling 
for the left-handed and right-handed chiralities 
is called an axial (a vector) gauge field.~\cite{bertlmann00}
In the presence of an axial component, 
the interaction between a gauge field and a fermion 
can be asymmetric for the two chiralities.
For example, in the case of the weak interactions for neutrinos,
only the left-handed chirality couples with a gauge field
and the theory is generally known as a chiral gauge theory.

A chiral gauge theory framework can be applied to graphene.
The energy band structure for the electrons 
in graphene~\cite{novoselov05,zhang05} 
has a structure similar to 
the massless fermion, 
in which the dynamics of the electrons
near the two Fermi points called the K and K$'$ points
in the two dimensional Brillouin zone
is governed by the Weyl equations.~\cite{wallace47}
Because the K and K$'$ points are related to each other under parity, 
two energy states near the K and K$'$ points
correspond to right-handed and left-handed chiralities,
respectively. 
The spin for a fermion corresponds to a pseudospin for graphene
which is expressed by a two-component wavefunction for the A and B
sublattices of a hexagonal lattice.~\cite{sasaki08ptps}
The corresponding pseudo-magnetic field for the pseudospin is given by an
axial gauge field that is induced by 
a deformation of the lattice in graphene.~\cite{kane97,sasaki08ptps,katsnelson08}  
The electronic properties of a graphene 
are thus described as a chiral gauge theory.~\cite{jackiw07}
An important point here is that 
the axial gauge field in graphene has 
different signs for the coupling constants about the two chiralities 
while the conventional electromagnetic (vector) gauge field does not.

\begin{figure}
 \begin{center}
  \includegraphics[scale=0.4]{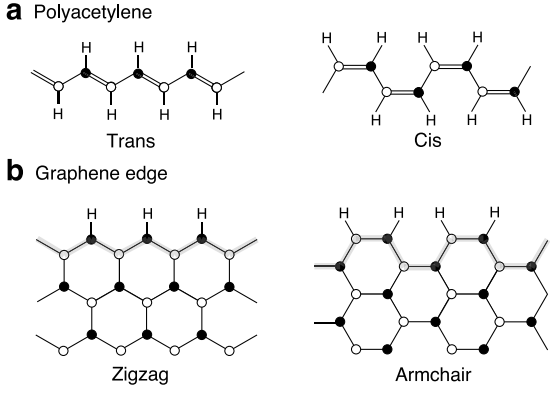}
 \end{center}
 \caption{Structures of a polyacetylene and a graphene edge.
 {\bf a}, Two possible isomers trans- and cis-polyacetylene.
 {\bf b}, Two principal edge structures: zigzag and armchair edges. 
 H denotes a hydrogen atom, and 
 carbon atoms are divided into A ($\bullet$) and B ($\circ$) atoms.
 }
 \label{fig:skelton}
\end{figure}

In a chiral gauge theory, 
the chiral symmetry breaking and the resultant mixing of chiralities are
of prime importance. 
In elementally particle physics, this symmetry breaking 
relates to the origin of the
mass of a fermion and the experimental investigations into the mass of
neutrinos are in progress. 
Since graphene is described by a chiral gauge theory,
a chirality mixing phenomenon in graphene is a matter of interest. 
In this paper, 
we show that a graphene nanoribbon which 
is a graphene with a finite width 
having two edges at the both sides,~\cite{jia09,jiao09,kosynkin09,chen07,stampfer09,gallagher10,han10} 
has a chirality mixed soliton solution
when applying strain to a graphene nanoribbon.
Two symmetric edge structures, that is, armchair and zigzag
edges are shown in Fig.~\ref{fig:skelton}.
It is known that the spatially localized electronic states, 
the edge states, appear near the zigzag edge.~\cite{tanaka87,fujita96,nakada96,son06,pereira06}
A chirality mixed soliton consists of 
two edge states belonging to different chiralities, 
and it is a natural extension of the concept of the topological soliton 
in trans-polyacetylene.~\cite{rice79,su79,takayama80,heeger88}

(Definition of gauge fields)

\begin{figure}
 \begin{center}
  \includegraphics[scale=0.4]{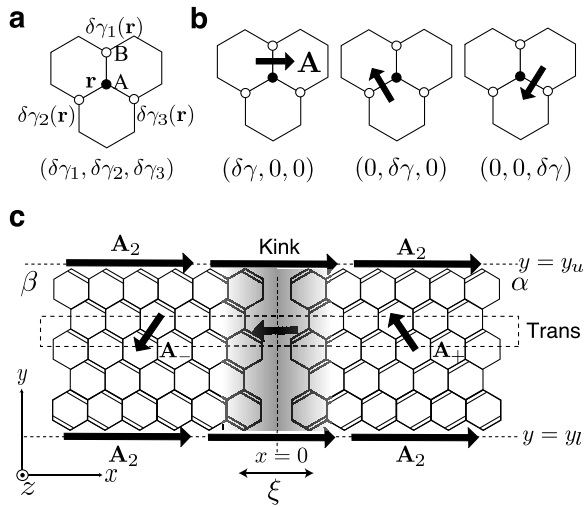}
 \end{center}
 \caption{Representing a lattice deformation in terms of the axial
 (deformation-induced) gauge field. 
 {\bf a}, A lattice deformation is defined by
 $(\delta\gamma_1,\delta\gamma_2,\delta\gamma_3)$. 
 {\bf b}, The direction of vector ${\bf A}$ is perpendicular to the lattice
 deformation. The directions of the arrows are for the case of a
 positive $\delta\gamma$.
 {\bf c}, The configuration of the axial gauge field ${\bf A}$
 for a trans zigzag nanoribbon.
 The two distinct bonding structures, $\alpha$ phase (${\bf A}_+$)
 and $\beta$ phase (${\bf A}_-$), 
 are combined together to form a a domain wall (a kink).
 The $y$-component of the field ${\bf A}_1(x)$ changes its sign 
 at $x=0$, which represents a kink structure. 
 The zigzag edges are represented by the field ${\bf A}_2(y)$.
 }
 \label{fig:gauge}
\end{figure}

First we review the chiral gauge theory of graphene.~\cite{sasaki08ptps}
A lattice deformation in graphene gives rise to 
a change of the nearest-neighbor hopping integral 
from the average value, $-\gamma$, as 
$-\gamma + \delta \gamma_{a}({\bf r})$,
where $a$ $(=1,2,3)$ denotes the direction of a bond
as shown in Fig.~\ref{fig:gauge}{\bf a}.
We define the axial gauge ${\bf A}({\bf r})=(A_x({\bf r}),A_y({\bf r}))$ 
by $\delta \gamma_{a}(\bf{r})$ as~\cite{kane97,sasaki08ptps,katsnelson08}
\begin{align}
 \begin{split}
  & v_{\rm F} A_x({\bf r}) = \delta \gamma_{1}({\bf r}) 
  - \frac{1}{2} \left\{
  \delta \gamma_{2}({\bf r})+ \delta \gamma_{3}({\bf r}) \right\}, \\
  & v_{\rm F} A_y({\bf r}) = \frac{\sqrt{3}}{2} 
  \left\{ \delta \gamma_{2}({\bf r})- \delta \gamma_{3}({\bf r}) \right\},
 \end{split}
 \label{eq:gauge}
\end{align}
where $v_{\rm F}$ is the Fermi velocity.
The direction of the vector ${\bf A}({\bf r})$ 
is perpendicular to that of the C-C bond 
with a modified hopping integral, 
as shown in Fig.~\ref{fig:gauge}{\bf b}.
The effective Hamiltonian 
for a deformed graphene 
is written by a 4$\times$4 matrix as~\cite{sasaki08ptps}
\begin{widetext}
\begin{align}
 {\hat H}\Psi({\bf r}) = v_{\rm F}
 \begin{pmatrix}
  \bsigma \cdot 
  \left( {\bf {\hat p}}+{\bf A}({\bf r})-e{\bf A}^{\rm em}({\bf r}) \right) 
  & \sigma_x \phi({\bf r}) \cr
  \sigma_x \phi^*({\bf r}) 
  &
  \bsigma' \cdot 
  \left( {\bf {\hat p}}-{\bf A}({\bf r})-e{\bf A}^{\rm em}({\bf r}) \right)
 \end{pmatrix}
 \begin{pmatrix}
  \Psi_{\rm K}({\bf r}) \cr \Psi_{\rm K'}({\bf r})
 \end{pmatrix},
 \label{eq:totH}
\end{align}
\end{widetext}
where the field $\phi({\bf r})$ relates to ${\bf A}({\bf r})$
as $v_{\rm F}\phi({\bf r})=(A_x({\bf r})+iA_y({\bf r}))e^{-2ik_{\rm F}x}$
in which $k_{\rm F}$ is the Fermi wave vector of the K point,
and ${\bf A}^{\rm em}({\bf r})$ is an electromagnetic gauge field.
Here, $\bsigma=(\sigma_x,\sigma_y)$ [$\bsigma'=(-\sigma_x,\sigma_y)$]
are the Pauli matrices which 
operate on the two-component spinors
$\Psi_{\rm K}({\bf r})$ and $\Psi_{\rm K'}({\bf r})$
for the pseudospin.
We use the units $v_{\rm F}=1$ and $\hbar = 1$, and thus
the momentum operator becomes ${\bf {\hat p}}=-i\nabla$.
A lattice deformation does not break time-reversal symmetry,
which appears as the different signs in front of the field 
${\bf A}({\bf r})$ for the two chiralities, while 
the electromagnetic gauge field ${\bf A}^{\rm em}({\bf r})$ breaks
time-reversal symmetry and has the same sign for the K and K$'$ points.
${\bf A}$ (${\bf A}^{\rm em}$) is an axial (a vector) gauge field.~\cite{bertlmann00}
Like the case of ${\bf A}^{\rm em}({\bf r})$, 
the field strength of ${\bf A}({\bf r})$ defined as 
$B_z({\bf r}) = \partial_x A_y({\bf r}) - \partial_y A_x({\bf r})$,
plays a fundamental role in discussing 
topological solitons and edge states, as we will show below.

It is straightforward to show using equation~(\ref{eq:gauge})
that the field $\phi$ behaves as a position-independent interaction
for the Kekul\'e distortion,~\cite{sasaki08ptps}
and then equation~(\ref{eq:totH}) is equivalent to the Dirac equation
with a mass $\phi$ 
in four-dimensional space-time without the $z$-component [$p_z = 0$] 
(see Appendix~\ref{app:1}).
Though the main concern of this paper is a chirality mixing due to a local mass
$\phi({\bf r})$, let us begin by considering the massless limit $\phi({\bf r})=0$
and examining the chirality eigenstate $\Psi_{\rm K}$ (right-handed
chirality) using the $2\times 2$ Hamiltonian,
$H({\bf r}) = \bsigma \cdot ({\bf {\hat p}}+{\bf A}({\bf r}))$.

(Topology of the gauge field)

In Fig.~\ref{fig:gauge}{\bf c}
the double bond represents a shrinking of the C-C bond
and the single bond denotes the absence of deformation. 
The $\alpha$ phase is defined 
as the bonding structure for the case of 
$(\delta\gamma_1,\delta\gamma_2,\delta\gamma_3)=(0,\delta\gamma,0)$,
while the $\beta$ phase is the case of 
$(\delta\gamma_1,\delta\gamma_2,\delta\gamma_3)=(0,0,\delta\gamma)$. 
From equation~(\ref{eq:gauge}), 
the corresponding ${\bf A}$ fields for the $\alpha$ and $\beta$ phases,
${\bf A}_+$ and ${\bf A}_-$, are given, respectively, by
${\bf A}_{\pm} = \left( -\delta\gamma/2,
 \pm \sqrt{3}\delta\gamma/2 \right)$.
For the skeleton of a trans-polyacetylene shown 
between the closed dashed lines of Fig.~\ref{fig:gauge}{\bf c}, 
it is well-known that a topological soliton appears 
when the configuration has a domain wall (a kink), that is, 
when the $\beta$ phase changes into the $\alpha$ phase at some position 
along the $x$-axis.~\cite{su79,takayama80,heeger88}
The gauge field for such a domain wall configuration 
for a zigzag nanoribbon is written as
\begin{align}
 {\bf A}_1(x)=(c_x,A_y(x)),
 \label{eq:A1}
\end{align}
where $c_x\equiv-\delta\gamma/2$,
$A_y(x) = -a_y$ ($a_y \equiv \sqrt{3}\delta\gamma/2$) when $x\ll-\xi$, 
and $A_y(x) = a_y$ when $x \gg \xi$.
Here, $\xi$ ($>0$) denotes the width of a kink (see Fig.~\ref{fig:gauge}{\bf c}).
In addition,
the gauge field which describes the edge structure
is given by ${\bf A}_2$. 
This ${\bf A}_2$ comes from the fact that 
the C-C bonds at the zigzag edge are cut.~\cite{sasaki06jpsj}
This cutting is represented by 
$(\delta\gamma_1,\delta\gamma_2,\delta\gamma_3)=(\gamma,0,0)$ 
at the edge, and ${\bf A}_2=\left( \gamma, 0 \right)$.
Since there are two zigzag edges at $y=y_u$ and $y=y_l$ 
in the zigzag nanoribbon (without a domain wall), 
${\bf A}_2(y)=(A_x(y),0)$ has a value only for 
$y=y_u$ and $y=y_l$ (the edge location), and otherwise $A_x(y)= 0$.
The total gauge field for a trans zigzag nanoribbon 
is given by the sum of 
${\bf A}_1(x)$ in equation~(\ref{eq:A1}) and 
${\bf A}_2(y)$ as
${\bf A}_1(x) + {\bf A}_2(y) =(c_x+A_x(y),A_y(x))$.
As a result, 
the (K point) Hamiltonian is given by
\begin{align}
 H({\bf r}) = 
 \sigma_x ({\hat p}_x+c_x+A_x(y)) + \sigma_y ({\hat p}_y+A_y(x)).
 \label{eq:HK}
\end{align}

(Zero-energy solution of $H({\bf r})$)

Here we assume that the energy eigenstates of $H({\bf r})$
in equation~(\ref{eq:HK}) has the form of 
$e^{ip_x x}\Psi_{p_x}(x,y) |\sigma \rangle$,
where $p_x$ is the quantum number 
and $|\sigma\rangle$ denotes the spinor eigenstate.
The energy eigenequation is rewritten as
\begin{align}
 & \left\{
 \sigma_x ({\hat p}_x+D_x+A_x(y)) + \sigma_y ({\hat p}_y+A_y(x))
 \right\} \Psi_{p_x}(x,y) |\sigma \rangle \nn \\
 & = E\Psi_{p_x}(x,y) |\sigma \rangle,
 \label{eq:maineq}
\end{align}
where $D_x\equiv p_x +c_x$.
We decompose this eigenequation into two parts 
by putting $\Psi_{p_x}(x,y) = \psi(x) \varphi(y)$ and $E=E_1+E_2$ 
as
\begin{align}
 & \left\{
  \sigma_x {\hat p}_x 
  + \sigma_y A_y(x) \right\} \psi(x) |\sigma \rangle= E_1
  \psi(x) |\sigma \rangle, \label{eq:Psec}\\
 & \left\{ 
  \sigma_x (D_x+A_x(y)) + 
  \sigma_y {\hat p}_y \right\} \varphi(y) |\sigma \rangle
  = E_2
  \varphi(y) |\sigma \rangle, \label{eq:Esec}
\end{align}
In general, the spinor eigenstate of the first equation
can not be identical to that of the second one.
However, in the special case that $E_1=E_2=0$,
the spinor eigenstates of these equations can 
be the same.
It is because that 
$H({\bf r})$ commutes with $\sigma_z$ for the zero-energy state,
$\left[H({\bf r}),\sigma_z\right]_- e^{ip_x x}\Psi^{E=0}_{p_x}({\bf
r})|\sigma \rangle=0$, and that 
the spinor eigenstate 
can be taken as the eigen spinor of $\sigma_z$
defined as $\sigma_z|\sigma_\pm\rangle = \pm |\sigma_\pm\rangle$, 
where
\begin{align*}
 |\sigma_+ \rangle =
 \begin{pmatrix}
  1 \cr 0
 \end{pmatrix},
  \ \
 |\sigma_- \rangle = 
 \begin{pmatrix}
  0 \cr 1
 \end{pmatrix}.
\end{align*}
Thus, the corresponding zero-energy states are pseudospin polarized
states, namely, the amplitude appears only one of the two sublattices.
In the following, we show that 
equations~(\ref{eq:Psec}) and (\ref{eq:Esec}) give, respectively, 
the topological soliton~\cite{su79,takayama80,heeger88,jackiw76} 
and the edge states.~\cite{sasaki06jpsj,sasaki08ptps}
From these zero-energy states for equations~(\ref{eq:Psec}) and
(\ref{eq:Esec}), a general zero-energy solution for equation~(\ref{eq:maineq})
can be constructed.

(Topological soliton)

Let us obtain the zero-energy 
soliton for equation~(\ref{eq:Psec}).
When $E_1=0$, the eigenequation is represented as
$\left\{ \sigma_x {\hat p}_x 
+ \sigma_y A_y(x) \right\} \psi_\pm(x) |\sigma_\pm \rangle=0$.
We have two solutions,
\begin{align}
 \psi_\pm(x) = N \exp \left( \pm \int^x A_y(x) dx \right),
 \label{eq:so}
\end{align}
where $N$ is a normalization constant.
When we use a trial function 
$A_y(x) = a_y \tanh(x/\xi)$, we get 
$\psi_\pm(x) = N \cosh^{\pm a_y \xi}(x/\xi)$.~\cite{rajaraman82}
Hence, when $a_y > 0$ (kink), only $\psi_-$ is selected,
while when $a_y < 0$ (anti-kink), only $\psi_+$ is selected.
The significance of a single zero-energy state is that
the particle-hole symmetric partner is given by itself,
which leads to the result 
that a soliton has no charge but has spin
$1/2$.~\cite{jackiw76,heeger88} 
The sign of $a_y$ corresponds to the sign of 
the field strength as 
$B_z(x) = a_y/(\xi \cosh^2 (x/\xi))$.
The sign of the $B_z$ field is essential
to a rule for obtaining the normalizable solution. 
This is easy to understand by noting that 
the square of $H({\bf r})$ is given by
$H({\bf r})^2=({\bf {\hat p}}+{\bf A}({\bf r}))^2+B_z({\bf r})\sigma_z$, 
which gives a positive coupling for $+B_z({\bf r})\sigma_z$.
Because $H({\bf r})^2 =0$ for a zero-energy state and 
$({\bf {\hat p}}+{\bf A}({\bf r}))^2$ is always a positive value, 
the zero-energy state needs to satisfy $+B_z({\bf r})\sigma_z<0$, so that 
a positive $B_z$ ($>0$) selects $|\sigma_-\rangle$ (or $\psi_-$) 
and a negative $B_z$ ($<0$) selects $|\sigma_+\rangle$ (or $\psi_+$).

(Edge states)

The derivation of the zero-energy edge states from equation~(\ref{eq:Esec}) 
is given in Ref.~\onlinecite{sasaki06jpsj}.
For the case of $D_x <0$,
there are degenerate zero-energy states given by 
\begin{align}
 \begin{split}
  & \varphi_+(y)|\sigma_+ \rangle = e^{D_x |y-y_u|} |\sigma_+ \rangle,
  \\
  & \varphi_-(y)|\sigma_- \rangle = e^{D_x |y-y_l|} |\sigma_- \rangle,
 \end{split}
 \label{eq:ed}
\end{align}
where $|D_x|^{-1}$ is the localization length.
As shown in Fig.~\ref{fig:gauge}{\bf c},
at the upper edge located at $y=y_u$, 
the $A_x(y)$ field increases abruptly
when $y$ approaches $y_u$ ($y \le y_u$).
Therefore, the corresponding $B_z$ field 
[$B_z(y) = - \partial_y A_x(y)$]
is pointing toward the negative $z$-axis there. 
Hence, only the $|\sigma_+\rangle$ state can appear near the upper edge.
In contrast, at the lower edge ($y=y_l$), 
the $A_x(y)$ field decreases abruptly as $y$ moves away from $y_l$
($y \ge y_l$). 
Therefore, the corresponding field strength is positive there, 
and only the $|\sigma_-\rangle$ state is selected near the lower edge.

(Soliton-edge state)

A zero-energy solution of equation~(\ref{eq:maineq}) 
is constructed by the product of the topological soliton 
$\psi_-(x)$ of equation~(\ref{eq:so}) 
and the edge state 
$\varphi_-(y) |\sigma_-\rangle$ of equation~(\ref{eq:ed}) as
\begin{align}
 \Psi^{-}_{p_x}(x,y) |\sigma_- \rangle = e^{- \int^x A_y(x) dx} 
 e^{D_x |y-y_l|} |\sigma_- \rangle.
 \label{eq:nsolp}
\end{align}
This new state is localized not only near the lower zigzag edge 
but also near the kink.
Since a kink satisfies $B_z(x)> 0$, 
this state $\Psi^{-}_{p_x}(x,y)|\sigma_-\rangle$ 
is the solution to equation~(\ref{eq:maineq}). 
If there is an anti-kink with $B_z(x)< 0$ at $x=0$, 
another zero-energy state given by
\begin{align}
 \Psi^{+}_{p_x}(x,y) |\sigma_+ \rangle = 
 e^{+ \int^x A_y(x) dx} e^{D_x |y-y_u|}
 |\sigma_+ \rangle,
 \label{eq:nsoln}
\end{align}
is the solution.
This state is localized near another zigzag edge 
and is also localized near the anti-kink.
In addition to these zero-energy solutions 
of the K point Hamiltonian,
there are zero-energy solutions
of the K$'$ point Hamiltonian.
Let the solutions for the K$'$ point be of the form of 
$\Psi_{-p_x}(x,y)|\sigma\rangle =\psi'(x)\varphi'(y)|\sigma\rangle$.
The energy eigenequation for the K$'$ point Hamiltonian,
$\bsigma' \cdot ({\bf {\hat p}}-{\bf A}({\bf r}))$, 
leads to a pair of energy eigenequations:
\begin{align*}
\begin{split}
 & \left\{
  \sigma_x {\hat p}_x 
  + \sigma_y A_y(x) \right\} \psi'(x) |\sigma \rangle= -E_1
  \psi'(x) |\sigma \rangle, \\
 & \left\{ 
  \sigma_x (D_x+A_x(y)) + 
  \sigma_y {\hat p}_y \right\} \varphi'(y) |\sigma \rangle
  = E_2
  \varphi'(y) |\sigma \rangle. 
\end{split}
\end{align*}
These eigenequations are the same as those 
given in equations~(\ref{eq:Psec}) and (\ref{eq:Esec})
(except for the unimportant sign change of $E_1$). 
As a result, the solutions to these equations 
are the same as equations~(\ref{eq:nsolp})
and (\ref{eq:nsoln}).
We thus have two zero-energy solutions
originating from the K and K$'$ points
for a given $p_x$.
This number of zero-energy states for a ribbon 
is different from a single zero-energy state
for a polyacetylene chain. 
This difference is attributed to the fact that 
the soliton for a polyacetylene chain
results from a chirality (or an intervalley) mixing.

(Chirality mixing)

\begin{figure}
 \begin{center}
  \includegraphics[scale=0.5]{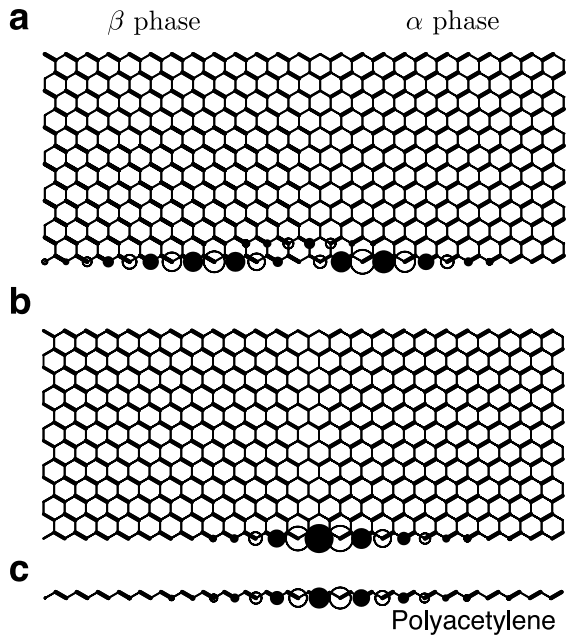}
 \end{center}
 \caption{Wave function patterns for zero-energy soliton.
 {\bf a}, An example of the wave function pattern. 
 The solid and empty circles represent 
 the phases ($+$ or $-$) of the wave function, 
 and the diameter of each circle is
 proportional to the amplitude. 
 We use $\delta\gamma=0.2\gamma$ and the kink profile of $\tanh(x/\xi)$
 with $\xi=2$\AA.
 {\bf b}, An example of the zero-energy soliton in a zigzag ribbon.
 Because the wave function of this example 
 appears only at the edge sites, this state 
 is identical to the topological soliton 
 in a trans-polyacetylene
 shown in {\bf c}.
 }
 \label{fig:wf}
\end{figure}

The zero-energy solutions 
given by equations~(\ref{eq:nsolp}) and (\ref{eq:nsoln}) 
have been obtained on the assumption that 
a chirality mixing between the K and K$'$ points 
can be neglected.
However, translational symmetry along the $x$-axis is broken
due to the presence of a kink (or anti-kink) and 
a kink itself causes a mixing of chiralities.
In this case, the eigenfunction may be written as a linear
combination of $\Psi^{\pm}_{p_x}$ for the K point 
and $\Psi^{\pm}_{-p_x}$ for the K$'$ point.
Their mixing is determined 
by the mass term which 
is expressed by means of valleyspin $\tau_a$
($a=0,1,2,3$) as 
$\sigma_x\left\{ \tau_1 {\rm Re}[\phi({\bf r})] -\tau_2 {\rm
Im}[\phi({\bf r})]\right\}$.
By putting 
$\Psi_{\rm K}= e^{-ikx}\tilde{\Psi}_{\rm K}$
and $\Psi_{\rm K'}= e^{+ikx}\tilde{\Psi}_{\rm K'}$
into equation~(\ref{eq:totH}) for a zigzag ribbon,
we obtain the equations for a general case:
\begin{align}
\begin{split}
 & \left\{\tau_3 \left(\sigma_x \hat{p}_x + \sigma_y A_y(x)\right)
 -  \tau_2 e^{i\tau_3 2\delta k x} \sigma_x A_y(x) \right\} \tilde{\Psi} = 
 E_1 \tilde{\Psi}, \\
 & \tau_0 \left\{ \sigma_x (c_x-k+A_x(y)) + \sigma_y \hat{p}_y \right\}
 \tilde{\Psi} = E_2 \tilde{\Psi},
\end{split}
\label{eq:geneH}
\end{align}
where $\delta k \equiv k_{\rm F}-k$.~\footnote{It is noted that we have neglected $\tau_1 \sigma_x (c_x + A_x(y))$
which would appear when we put ${\bf A}_1(x) + {\bf A}_2(y)$ 
into the definition of the field $\phi({\bf r})$. 
It is because 
a constant field $c_x$ and the zigzag edge $A_x(y)$
are irrelevant to a chirality scattering
process.~\cite{sasaki10-forward}}
The last term on the left-hand side of the first equation 
of equation~(\ref{eq:geneH})
shows that the effective domain wall profile
for the mixing term
is an oscillating function of $x$, 
in contrast to a smooth function 
of the intravalley mixing term for the second term.
It is straightforward to find a zero-energy solution of
equation~(\ref{eq:geneH}) as
$\tilde{\Psi}^\pm = U(x) \psi_\pm(x) \varphi_\pm(y)
 |\sigma_\pm \rangle$,
where $U(x)$ is a matrix for valleyspin
which satisfies the equation
$\partial_x U(x)-\tau_1 e^{i\tau_3 2\delta k x}A_y(x) U(x)=0$.
Due to the chirality mixing,
the actual wave function of a zero-energy state
in a zigzag nanoribbon can be complicated.
One example of the wave function is shown in Fig.~\ref{fig:wf}{\bf a}.

(Soliton in polyacetylene)

Equation (\ref{eq:geneH}) can be solved analytically 
for the special case of $\delta k=0$ ($k=k_{\rm F}$).
In this case, we obtain simultaneous differential equations:
\begin{align}
\begin{split}
 & \sigma_x\left\{\frac{\tau_3}{2} \hat{p}_x-\tau_2A_y(x)\right\} \tilde{\Psi} =
 E_1 \tilde{\Psi},
 \\
 &
 \tau_3 \left\{\frac{\sigma_x}{2} \hat{p}_x+\sigma_y A_y(x)\right\}
 \tilde{\Psi} = E_2 \tilde{\Psi}, 
 \\
 & \tau_0 \left\{ \sigma_x (c_x-k_{\rm F}+A_x(y)) + \sigma_y \hat{p}_y \right\}
 \tilde{\Psi} = E_3 \tilde{\Psi}.
\end{split}
\label{eq:conpoly}
\end{align}
The first term gives rise to a chirality mixing.
For a zero-energy solution of the first term, 
the spinor eigenstate should be the eigen spinor of $\tau_1$ defined as 
$\tau_1|\tau_\pm \rangle = \pm|\tau_\pm \rangle$,
which shows a strong chirality mixing. 
The zero-energy solutions for the first two terms 
are given by $\tilde{\psi}_\pm(x)|\sigma_\pm\rangle \otimes |\tau_\mp \rangle$,
where $\tilde{\psi}_\pm(x) = N \exp\left( \pm \int^x 2A_y(x)dx \right)$.
This state is a valleyspin unpolarized state and 
also a pseudospin polarized state, and 
these properties are consistent with 
those of the topological soliton in polyacetylene.~\cite{heeger83}
The third equation in equation~(\ref{eq:conpoly})
describes the edge state having the shortest 
localization length 
since the localization length is given by $|c_x-k_{\rm F}|^{-1}$ 
which vanishes in the continuum limit.
The resulting zero-energy solution 
of equation (\ref{eq:conpoly}) given by
$\tilde{\Psi}({\bf
r})=\tilde{\psi}_\pm(x)\varphi_\pm(y)|\sigma_\pm\rangle \otimes
|\tau_\mp \rangle$ 
corresponds to Fig.~\ref{fig:wf}{\bf b} which 
reproduces a topological soliton in polyacetylene 
at the zigzag edge sites (see Fig.~\ref{fig:wf}{\bf c} for comparison).
Note that the soliton can move along the zigzag edge and the soliton has
a mass.
In the case of polyacetylene, the soliton mass is estimated to be around 
$6m_e$, where $m_e$ is the mass of the free electron.~\cite{heeger83} 
In the case of the ribbon, we obtain $65(W/\xi) m_e$, where $W$ denotes 
the ribbon width. This result reproduces the soliton mass in polyacetylene
when $W=a$ and $\xi=10a$, where $a$ is the lattice constant.

To further elucidate the effect of the edge on the soliton,
we consider the solitons of an armchair tube, a metallic zigzag tube, 
and a metallic armchair ribbon in Appendices~\ref{app:4} and~\ref{app:5}.
We show that the chirality mixing is negligible for the zero-energy
states in these tubes.
A metallic armchair ribbon produces chirality mixed solitons 
when there is a domain wall. 
The solitons are not localized near the edge since there is no edge
states near the armchair edge.
This feature is in contrast to that of the soliton in a zigzag nanoribbon.
See Appendices~\ref{app:4} and~\ref{app:5} for more details.

(Discussion)

We can use equation~(\ref{eq:gauge}) 
for a lattice deformation induced by a strain.
Let ${\bf u}({\bf r})=(u_x({\bf r}),y_y({\bf r}))$ 
be the displacement vector
of a carbon atom at ${\bf r}$, 
the axial gauge field is written as~\cite{suzuura02,guinea10,sasaki10-physicaE} 
\begin{align*}
 \begin{split}
  & A_x({\bf r}) = g \left[ - \frac{\partial u_x({\bf r})}{\partial x} +
  \frac{\partial u_y({\bf r})}{\partial y} \right], \\
  & A_y({\bf r}) = g \left[ \frac{\partial u_x({\bf r})}{\partial y} +
  \frac{\partial u_y({\bf r})}{\partial x} \right], 
 \end{split}
\end{align*}
where $g$ is the electron-phonon coupling.
An interesting consequence of this is that 
the field configurations which are equivalent to ${\bf A}_\pm$
may be realized when an appropriate strain is applied to a sample.
For example, 
a ``${\rm V}$'' shape graphene nanoribbon caused by 
an acoustic shear deformation given by $u_x=0$ and 
$u_y(x) = u \ln [\cosh(x/\xi)]$ with $u=\xi(a_y/g)$, 
can reproduce the gauge field 
representing a bond alternation (a domain wall) 
in zigzag ribbons.
Because $g \simeq \gamma$,~\cite{sasaki10-physicaE}
$u$ is smaller than $\xi$ by a factor of $\delta\gamma/\gamma$.
This shows that a domain wall can be realized by a strain of
$\sim$ 10\%.~\cite{guinea10} 
Similarly, a strain produces a localized soliton
in a metallic armchair nanoribbon (see Appendix~\ref{app:5}).
On the other hand, the ribbon does not support 
the edge states without the strain.
A pseudospin polarized wavefunction pattern 
that is spatially localized 
near the bottom of a ``${\rm V}$'' shape graphene nanoribbon 
is the indication of a chirality mixed soliton.
Note that a strain makes it possible to observe 
the soliton by means of a scanning tunneling microscopy (STM)
experiment, in contrast to that 
the STM is unable to detect a soliton in polyacetylene 
since the soliton is moving.
Moreover, it was suggested recently by Guinea {\it et al.}~\cite{guinea10} 
that a uniform $B_z$ field may be realized in graphene by 
a strain-induced lattice deformation, 
which is an interesting consequence.
If this is the case, it is expected that 
the Landau level appears only for one chirality and 
the other chirality decouples from the gauge fields
in the presence of a magnetic field which eliminates $B_z$ for one
chirality.
Then the chiral symmetry in graphene is maximally broken, and 
this situation is similar to the case of weak interactions 
in elementally particle physics.

\section*{Acknowledgments}

K.S, K.W, and T.E are supported by 
a Grant-in-Aid for Specially Promoted Research
(No.~20001006) from the Ministry of Education, Culture, Sports, Science
and Technology (MEXT).
R.S acknowledges a MEXT Grant (No.~20241023).
M.S.D acknowledges grant NSF/DMR 07-04197.
K.S. thanks Professor 
Francisco (Paco) Guinea for useful comments. 

\appendix

\section{Original Dirac Hamiltonian}\label{app:1}

The original Dirac Hamiltonian is written as 
\begin{widetext}
\begin{align*}
 {\hat H}\Psi({\bf r}) = 
 \begin{pmatrix}
  \bsigma \cdot 
  \left( {\bf {\hat p}}+ {\bf {\cal A}}({\bf r}) 
  +{\bf {\cal V}}({\bf r}) \right) 
  & m \cr
  m &
  - \bsigma \cdot 
  \left( {\bf {\hat p}}- {\bf {\cal A}}({\bf r}) 
  +{\bf {\cal V}}({\bf r})\right)
 \end{pmatrix}
 \begin{pmatrix}
  \Psi_{\rm R}({\bf r}) \cr \Psi_{\rm L}({\bf r})
 \end{pmatrix},
\end{align*}
\end{widetext}
where $m$ is the mass of fermion.
The electronic Hamiltonian for graphene 
corresponds to the case in which 
$\Psi_{\rm R}\to \Psi_{\rm K}$,
$\Psi_{\rm L}\to \sigma_x \Psi_{\rm K'}$, and 
$m \to \phi({\bf r})$.
The vector gauge field ${\bf {\cal V}}({\bf r})$
and axial gauge field ${\bf {\cal A}}({\bf r})$
correspond to $e{\bf A}^{\rm em}({\bf r})$ 
and ${\bf A}({\bf r})$, respectively.
The third component such as ${\hat p}_z$ 
is assumed to be zero when we identify the 
original Dirac equation (in $3+1$ dimensional space-time)
with the effective Hamiltonian for graphene 
(in $2+1$ dimensional space-time).

\section{Solitons in armchair nanotubes}\label{app:4}

Here we consider the solitons in armchair nanotubes.
The K point Hamiltonian is given by 
removing $A_x(y)$ from equation~(5). 
By putting $\Psi_{p_x}(x,y)= e^{-iD_x x} \psi(x) e^{ip_y y}$ 
into the energy eigenequation~(5), we obtain 
$\left\{ \partial_x \mp (p_y + A_y(x)) \right\} \psi_\pm(x) = 0$
for the zero-energy state. 
It follows that the function $\psi_\pm(x)$ 
contains the exponential function $\exp(\pm p_y x)$,
so that either $\psi_+(x)$ with $p_y=0$ or $\psi_-(x)$ with $p_y=0$ 
can be a normalizable solution.
The momentum $p_y$ is quantized by a periodic boundary condition 
around the tube's axis, 
and a zero-momentum state $p_y =0$ satisfies the boundary condition 
for any armchair nanotube.~\cite{saito92apl}
The solution with $p_y=0$ is a topological soliton.
From equation~(2), we obtain 
the chirality mixing term as
$\sigma_x\left\{ \tau_1 {\rm Re}[\phi(x)] -\tau_2 {\rm Im}[\phi(x)]\right\}$,
where $\phi(x)=iA_y(x)e^{-2ik_{\rm F}x}$.
This mixing term is small because 
a smooth function $A_y(x)$ of $x$ 
is multiplied by a rapid oscillating function $e^{-2ik_{\rm F}x}$.
Moreover, the chirality mixing term does not cause
a first order energy shift 
since the unperturbed states $\psi_\pm(x)|\sigma_\pm \rangle$ 
are pseudospin polarized states 
satisfying $\langle \sigma_\pm| \sigma_x |\sigma_\pm\rangle=0$.
For these reasons the chirality mixing is negligible
in the case of an armchair nanotube.

Note that the zero-energy solitons in an armchair nanotube 
obtained above are distinct from the topological soliton in
polyacetylene. 
The chirality mixing term is irrelevant to the solitons
in armchair nanotubes, while it is relevant to the topological soliton
in polyacetylene.

\section{Solitons in zigzag nanotubes and armchair ribbons}\label{app:5}

Let us examine solitons in a zigzag nanotube and an armchair ribbon.
The existence of a zero-energy topological soliton in a zigzag tube 
requires two factors: 
the ${\bf A}$ field topology and the presence of the Dirac singularity. 
The ${\bf A}$ field topology can be understood 
by noting that the basic unit of structure
is a cis-polyacetylene for which the two phases 
shown in Fig.~\ref{fig:wf-arm}{\bf a}
can be considered.~\cite{heeger88}
The $\alpha$ phase is defined by 
$(\delta\gamma_1,\delta\gamma_2,\delta\gamma_3)=(0,\delta\gamma,\delta\gamma)$,
and the $\beta$ phase is 
$(\delta\gamma_1,\delta\gamma_2,\delta\gamma_3)=(\delta\gamma,0,0)$.
From equation~(1), 
the corresponding gauge fields for the $\alpha$ and $\beta$ phases,
${\bf A}_+$ and ${\bf A}_-$, are given, respectively, by
${\bf A}_{\pm} = \left( \mp \delta\gamma, 0 \right)$
[see Fig.~\ref{fig:wf-arm}{\bf b}].
A domain wall kink is represented by 
${\bf A}_1(y)=(A_x(y),0)$ with $A_x(y)=-\delta\gamma \tanh(y/\xi)$.
By assuming that the wave function is of the form of 
$e^{ip_y y}\psi(x)\varphi(y) |\sigma \rangle$,
we have a pair of the eigenequations from the K point Hamiltonian as
\begin{align*}
\begin{split}
 & \left\{
  \sigma_x A_x(y) + \sigma_y {\hat p}_y \right\} \varphi(y) |\sigma \rangle= E_1
  \varphi(y) |\sigma \rangle, \\
 & \left\{ 
  \sigma_x {\hat p}_x + 
  \sigma_y p_y \right\} \psi(x) |\sigma \rangle = E_2 \psi(x) |\sigma
 \rangle.
\end{split}
\end{align*}
The first equation 
possesses a zero-energy topological soliton.
Therefore, when there is a zero-energy state for the second equation,
the K point Hamiltonian may possess a mixed zero-energy solution.
The state with $p_x=0$ and $p_y=0$, i.e., 
the state at the Dirac singularity, 
can satisfy the second equation with $E_2=0$.
Since $p_x$ is quantized by a periodic boundary condition 
around the tube's axis, this state with vanishing wave vector 
exists only for ``metallic'' zigzag tubes.~\cite{saito92apl}
For ``semiconducting'' zigzag tubes, the quantized $p_x$ misses the
Dirac singularity, and therefore such a zero-energy topological soliton
does not exist. 
Thus, only the presence of a non-vanishing $B_z$ field strength
does not necessarily result in the presence of a zero-energy state.
In addition to a domain wall, the Dirac singularity 
is rather essential for the presence of a zero-energy state.
Note that a non-topological excitation, a polaron, may exist
even in ``semiconducting'' zigzag tubes
near a bound kink-antikink pair.~\cite{heeger88}

\begin{figure}
 \begin{center}
  \includegraphics[scale=0.5]{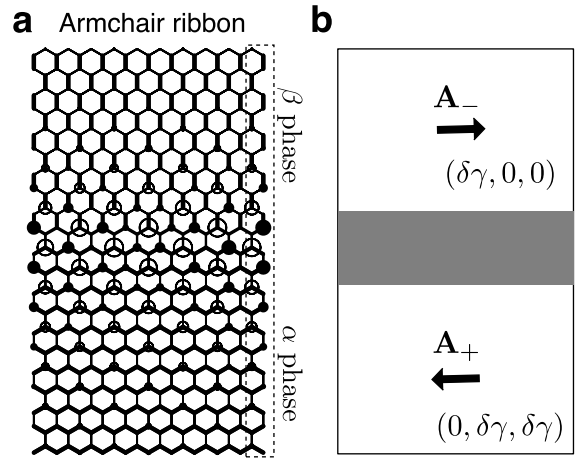}
 \end{center}
 \caption{Soliton in an armchair nanoribbon.
 {\bf a}, The wave function of a topological soliton in a
 metallic armchair nanoribbon obtained from a tight-binding model. 
 {\bf b}, The two phases ${\bf A}_+$ and ${\bf A}_-$ are 
 separated by a domain wall kink distortion represented by the shaded region.\label{fig:wf-arm}
 }
 \label{fig:wf-arm}
\end{figure}

The localization pattern of a topological soliton 
is sensitive to the lattice structure of the edge of a nanoribbon.
To illustrate this, we show 
the wave function of a topological soliton in a ``metallic'' armchair
nanoribbon in Fig.~\ref{fig:wf-arm}{\bf a}. 
The soliton is extended along the kink, which is contrasted with 
the localized feature of the wave function of a zero-energy state
in a zigzag nanoribbon shown in Fig.~3{\bf a} (in the text). 
This difference is a consequence of the fact that 
unrolling a zigzag tube can be represented 
by a strong intervalley mixing term $\phi({\bf r})$ 
at the armchair edge,
and that this field $\phi({\bf r})$ 
does not destroy the Dirac singularity.~\cite{sasaki10-chiral}
As a result, a topological soliton appears in a ``metallic'' armchair
ribbon, as illustrated in Fig.~\ref{fig:wf-arm}{\bf a}.
It is interesting to note that 
unrolling a ``metallic'' zigzag tube does result in
a ``semiconducting'' armchair ribbon.
This implies that a topological soliton in a ``metallic'' zigzag tube
disappears when the tube is unrolled since the Dirac singularity 
also disappears then.


\end{document}